\documentclass[12pt]{article}
\usepackage{amssymb}
\usepackage{graphicx}
\usepackage{amsmath}
\usepackage{enumerate}
\usepackage[english]{babel}
\usepackage{latexsym}

\def\sod{{\leavevmode\setbox1=\hbox{d}%
\hbox to 1.05\wd1{d\kern-0.4ex{\char039}\hss}}}
\def\sot{{\leavevmode\setbox1=\hbox{t}%
\hbox to \wd1{t\kern-0.5ex{\char039}\hss}}}
\def\sol{l\kern-0.3ex\raise0.1ex\hbox{'}\kern-0.10ex}
\def\soL{L\kern-0.8ex\raise0.1ex\hbox{'}\kern0.1ex}
\begin{document}
\title{Study of anomalous kinetics of the annihilation reaction $\textit{A} +\textit{A}  \rightarrow \varnothing$}

\author{M. Hnatich$^{1,2}$, J. Honkonen$^{3}$, T. Lu\v{c}ivjansk\'y$^{1,2}$}

\maketitle
 $^{1}$ Institute of Experimental
Physics, Slovak Academy of Sciences, Watsonova 47, 040 01
Ko\v{s}ice, Slovakia,\\
$^{2}$ Faculty of Sciences, P.J. \v{S}afarik
University, Moyzesova 16, 040 01 Ko\v{s}ice, Slovakia,\\
$^3$
Department of Military Technology, National Defence University,
P.O.~Box~7, 00861, Helsinki, Finland.\\
\begin{abstract}
 Using perturbative renormalization group we study the influence of
random velocity field on the kinetics of single-species annihilation
reaction $\textit{A} +\textit{A} \rightarrow \varnothing$ at and
below its critical dimension $d_c=2$. We use the second-quantization
formalism of Doi to cast the stochastic problem into a
field-theoretic form. The reaction is analyzed near two dimensions
by means of two-parameter expansion in $\epsilon,\Delta$, where
$\epsilon$ is the deviation from the Kolmogorov scaling and $\Delta$
is the deviation from the space dimension $2$. All of the relevant
quantities are evaluated to the second order of the perturbation
scheme and long-time asymptotic
 behaviour of stochastic system under
consideration is analyzed.
\end{abstract}


\section{Introduction}
The annihilation reaction $\textit{A} +\textit{A} \rightarrow
\varnothing$, where the reactans are influenced by some external
advective field, are  one of the simplest examples of the nonlinear
statistical systems. This type of reactions may be observed in
diverse chemical, biological or physical systems. Pollutants in the
atmosphere of the Earth is typical example of such stochastic
system. For this type of reaction in low space dimensions the usual
description by means of kinetic rate equation is not sufficient and
the effect of density fluctuations must be taken into the account.
  The origin of this effect could be tracked
down to the physics of diffusion  \cite{Lee,Cardy}.  \\
A theoretical analysis of this reaction has been carried out for the
case of time-independent random drift \cite{Deem}. However, for a
more realistic description of the effects of the velocity
fluctuations a time-dependent random drift would be more
appropriate. The most suitable approach to this end is based on the
use of the stochastic Navier-Stokes equations
 \cite{Adzhemyan}.
To gain an insight into the problem, it is useful to consider for
the velocity fluctuations a simplified model less complicated
calculationally. In this work the Kraichnan model with finite
correlation time \cite{Adzhemyan2} is used for modelling the
advective velocity field. The motivation for using it is not only
its simplicity, but this model also reproduces many of the observed
phenomena in the genuine turbulence mass transfer \cite{Frisch}. In
this model there are two special limits: the self-similar
white-in-time velocity field (''rapid-change mode'') and the
time-independent (quenched) velocity
field. Both of them are considered in this paper.\\
In this paper the long-time behaviour of the annihilation reaction
$\textit{A} +\textit{A} \rightarrow \varnothing$ is studied in the
presence of a synthetic velocity field at and below its critical
dimension $d_c=2$. The second-quantization  formalism of Doi
\cite{Doi} is used to cast the reaction part into a field-theoretic
form.
 It is assumed that
the statistics of velocity field is Gaussian with finite correlation
time. The theoretical model for such a reaction is constructed and
with the use of the perturbative renormalization group possible IR
regimes are calculated to the second order.
\section{Field-theoretic model}
A classical reaction system may be cast into the language of
second-quantization with the use of the creation and annihilation
operators $\psi$ and $\psi^\dagger$ and the vacuum state $|0\rangle$
\cite{Doi}
satisfying the usual commutation relations, \\
\begin{eqnarray}
 & &[\psi(\bold{x}),\psi^\dagger(\bold{x^{'}})]=\delta(\bold{x}-\bold{x^{'}})\nonumber\\
 & &[\psi(\bold{x}),\psi(\bold{x^{'}})]=[\psi^\dagger(\bold{x}),\psi^\dagger(\bold{x^{'}})] = 0 \nonumber\\
 & &\psi(\bold{x})|0\rangle   = 0, \langle 0| \psi^\dagger(\bold{x}) = 0, \langle 0|0 \rangle = 1
\end{eqnarray}
Let $P(\{n_i\},t)$ represent the joint probability distribution function for observing $n_i$ particles at positions
$\mathbf{x}_i$, then the state vector of the classical many-particle system is defined as the sum over
all occupation numbers
\begin{eqnarray}
 | \Phi(t)\rangle = \sum_{n_{i}} P(\{ n_i \},t) | \{ n_i \} \rangle,
\end{eqnarray}
where the basic vectors are conventionally defined as
\begin{eqnarray}
 | \{ n_i \} \rangle = \prod_{i} [\psi^\dagger(\bold{x}_i) ]^{n_i} |0\rangle.
\end{eqnarray}
Now the whole set of coupled equations for the probability density
functions can be rewritten in the compact form \cite{Lee,Doi}
\begin{eqnarray}
 \frac{\partial}{\partial t}| \Phi(t)\rangle = -\hat{H} | \Phi(t) \rangle,
\end{eqnarray}
where $\hat{H} = \hat{H}_A+\hat{H}_D+\hat{H}_R$. It can be shown
\cite{Hnatic} that for the annihilation reaction
studied these terms are expressed as
\begin{eqnarray}
   & &\hat{H}_A = \int d\bold{x} \psi^\dagger \nabla [\bold{v}(\bold{x},t) \psi(\bold{x})],\nonumber\\
   & &\hat{H}_D = - D_{0} \int d\bold{x} \psi^\dagger \nabla^{2} \psi(\bold{x}),\nonumber\\
   & &\hat{H}_R = \lambda_{0}D_0 \int d\bold{x} (\psi^\dagger)^{2} \psi^{2},
   \label{eqs:hamil_react}
\end{eqnarray}
corresponding to the advection, diffusion and reaction part. Because of dimensional reasons we have extracted
the diffusion constant $D_0$ from the rate constant $K = \lambda_0 D_0$.
Mean values of physical quantities may be expressed \cite{Hnatic} via the relation
\begin{eqnarray}
  \langle A(t) \rangle = \langle 0| T\biggl(A\{[\psi^\dagger(t)+1]\psi(t) \} \exp(-\int^{\infty}_{0} \hat{H}'_{I} dt + n_{0} \int d\bold{x} \psi^\dagger(\bold{x},0) ) \biggl)|0\rangle
  \label{eq:aver2}
\end{eqnarray}
The expectation value of time-ordered product (\ref{eq:aver2}) may
be cast by the standard operation \cite{Vasiliev} into the form of a
functional integral over complex-conjugate scalar fields
$\psi^\dagger(\mathbf{x},t)$ and $\psi(\mathbf{x},t)$:
\begin{eqnarray}
 \langle A(t) \rangle = \int \mathcal{D}\psi^\dagger \mathcal{D}\psi A\{ [\psi^\dagger(t)+1]\psi(t)\} \rm{e}^{S_1}
  \label{expect_val}
\end{eqnarray}
where the unrenormalized action $S_{1}$ for the annihilation
reaction $\textit{A} +\textit{A} \rightarrow \varnothing$  is
\begin{eqnarray}
  S_{1} & =  &- \int^{\infty}_{0} dt \int d\bold{x} \{ \psi^\dagger \partial_{t} \psi + \psi^\dagger\nabla(\bold{v}\psi) -D_{0}\psi^\dagger\nabla^{2}\psi
  + \nonumber\\
  & & \lambda_{0} D_{0}[2\psi^\dagger+(\psi^\dagger)^{2}]\psi^{2} +
  n_{0}\int d\bold{x} \psi^\dagger(\bold{x},0)  \}.
  \label{eq:S_1}
\end{eqnarray}
The most realistic description of the velocity field ${\bf v}(x)$ is
based on the use of stochastic Navier-Stokes equation. However, in
this paper we shall study a simplified model in which we prescribe
statistical properties of the velocity field. Let assume that ${\bf
v}(x)$ is a random Gaussian variable with zero mean and the
correlator
\begin{eqnarray}
 \langle v_i(x) v_j(x') \rangle = \int \frac{d{\bf k} d\omega}{(2\pi)^{d+1}} P_{ij}({\bf k}) D_v(\omega,{\bf k})
  \exp[-i\omega(t-t') + i{\bf k}.({\bf x}-{\bf x'})].
\label{correlator_v}
\end{eqnarray}
Here $P_{ij}({\bf k}) = \delta_{ij}-k_ik_j/k^2$ is the transverse
projection operator, $k=|{\bf k}|$ is the wave number and the kernel
function $D_v$ is assumed to have the following form
\begin{eqnarray}
  D_{v}(\omega,{\bf k}) & = & \frac{g_0 D_0^3 k^{2-2\Delta-2\epsilon}}{\omega^2+u_0^2 D_0^2 (k^{2-\eta})^2}.
  \label{correlator_v_fourier}
\end{eqnarray}
Here $g_0$ is the coupling constant (small parameter of the ordinary
perturbation theory) and the exponents $\epsilon$,$\Delta$ and
$\eta$ play the role of small expansion parameters. They could be
regarded as an analog of the expansion parameter $\epsilon=4-d$ in
the usual sense of dimensional regularization. However, in this
paper $\epsilon$ should be understood as deviation of exponent of
the power law from that of the Kolmogorov scaling \cite{Frisch},
whereas $\Delta$ is defined as the deviation from the space
dimension two via relation $d=2+2\Delta$. The exponent $\eta$ is
related to the reciprocal of the correlation time at the wave number
$k$. The parameter $u_0$ may be used for labelling of the fixed
points and has the meaning of the ratio of velocity correlation time
and the scalar turnover time \cite{Adzhemyan2}. Although in the real
calculations $\epsilon$ is treated as a small parameter, the real
problem corresponds to the value $\epsilon=4/3$.
\\
 It is interesting to note that the model for the advection field ${\bf v}(x)$ contains two cases of special interest:
\begin{enumerate}[(a)]
 \item in the limit $u_0 \rightarrow \infty, g_0' \equiv g_0/u_0^2 = const$ we get the 'the rapid-change model'
  $ D_v(\omega,{\bf k}) \rightarrow g_0' D_0 k^{-2-2\Delta-2\epsilon+\eta}$,
  which is characterized by the white-in-time nature of the velocity correlator.
 \item limit $u_0 \rightarrow 0, g_0''\equiv g_0/u_0 = const$ corresponds to the case of a frozen velocity field
  $ D_{v}(\omega,{\bf k}) \rightarrow g_0'' D_0^2 \pi \delta(\omega) k^{2\Delta-2\epsilon}$,
 when the velocity field is quenched (time-independent).
\end{enumerate}
The averaging procedure with respect to the velocity field ${\bf v}(x)$ 
may be performed with the aid of the following action functional
\begin{eqnarray}
  S_2 = -\frac{1}{2}\int dt\mbox{ }dt'\mbox{ }d{\bf x}\mbox{ }d{\bf x'} \mbox{ } {\bf v}(t,{\bf x}) D_v^{-1}(t-t',{\bf x}-{\bf x'}) {\bf v}(t',{\bf x'}),
  \label{eq:S_2}
\end{eqnarray}
where $D_v^{-1}$ is the inverse correlator (\ref{correlator_v}) (in
the sense of the Fourier transform). The expectation value of any
relevant physical observable may be calculated using the complete
weight functional $\mathcal{W}(\psi^\dagger,\psi,{\bf v}) = {\rm
e}^{S_1+S_2}$, where $S_1$ and $S_2$ are the action functionals
(\ref{eq:S_1}) and (\ref{eq:S_2}).
\section{Power counting and UV renormalization}
In order to apply the minimal subtraction scheme for the evaluation 
of renormalization constants, an analysis of possible superficial
divergences has to be performed. 
For the power counting in the actions (\ref{eq:S_1}) and
(\ref{eq:S_2}) we use the scheme \cite{Adzhemyan}, in which to each
quantity $Q$ two canonical dimensions are assigned, one with respect
to the wave number $d_Q^k$ and the other to the frequency
$d_Q^\omega$. The normalization for these dimensions is
conventional: $d^\omega_\omega=-d_t^\omega=1,d_k^k=-d_x^k=1$ and
$d_k^\omega=d^k_\omega=0$. The canonical (engineering)  dimensions
for fields and parameters are derived from the condition for action
to be a dimensionless quantity.\\
The quadratic part of the action (\ref{eq:S_1}) determines only the
canonical dimensions of the product of fields $\psi^\dagger\psi$. In
order to keep both terms in the nonlinear part
$$
\lambda_0 D_0 \int
dtd\mathbf{x}[2\psi^\dagger+(\psi^\dagger)^2]\psi^2
$$
of the action, the field $\psi^\dagger$ must be dimensionless. If
the field $\psi^\dagger$ has a positive canonical dimension, then
the quartic term should be discarded as irrelevant by power
counting. The action with cubic term only, however, does not
generate any loop integrals
corresponding to density fluctuations and thus is uninteresting for the analysis of fluctuation effects.\\
Using this choice we get the following canonical dimensions for fields and parameters in $d$-dimensional space:\\
\begin{center}
\begin{tabular}{|c|c|c|c|c|c|c|c|}
  \hline
  $Q$ & $\psi$ & $\psi^\dagger$ & $v$ & $D_0$ & $u_0$ & $\lambda$ & $g_0$ \\ \hline
  $d_Q^k$ & $d$ & $0$ & $-1$ & $-2$ & $\eta$ & $2-d$ & $2\epsilon+\eta$ \\ \hline
  $d_Q^\omega $& $0 $ & $0$ & $1$ & $1$ & $0$ & $0$ & $0$ \\ \hline
  $d_Q $& $d$ & $0$ & $1$ & $0$ & $\eta$ & $2-d$ & $2\epsilon+\eta$ \\ \hline
\end{tabular}
\end{center}
Here, $d_Q=2d_Q^\omega+d^k_Q$ is the total canonical dimension and it is determined from the condition that the parabolic
differential operator of the diffusion scales uniformly under the transformation
$k\rightarrow \mu k, \omega\rightarrow \mu^2\omega$.

The model is logarithmic for $\epsilon=\Delta=0$, the ultraviolet (UV) divergences have the form of poles in various
linear combinations of $\epsilon$ and $\Delta$. The total canonical dimension of an arbitrary one-particle irreducible Green (1PI)
function $\Gamma=\langle\Phi\ldots \Phi  \rangle_{\rm 1-ir} $ is given by relation
\begin{eqnarray}
  d_\Gamma= d + 2 - N_\Phi d_\Phi,
\end{eqnarray}
where $N_\Phi=\{N_{\psi^\dagger},N_\psi,N_v \}$ are the numbers of corresponding external fields. Superficial UV divergences may be present
only in those $\Gamma$ functions for which $d_\Gamma$ is a non-negative integer.
Although the canonical dimension of the field $\psi^\dagger$ is zero, there is no proliferation of superficial divergent
graphs with arbitrary number of external $\psi^\dagger$ legs. This is due to the fact that $n_{\psi^\dagger}\leq n_\psi$, which
is easily seen by a straightforward analysis of the graphs \cite{Lee}.
As has already been shown \cite{Adzhemyan2} that the divergences in (1PI) Green functions containing at least one velocity field ${\bf v}$
could be removed by the only counterterm of the form $\psi^\dagger \partial^2 \psi$, which
leads to the following renormalization of parameters $g_0, D_0$ and $u_0$:
\begin{eqnarray}
   D_0 = D Z_D,\hskip1cm g_0=g\mu^{2\epsilon+\eta}Z_g,\hskip1cm u_0=u\mu^\eta Z_u,
   \label{eq:ren_rel1}
\end{eqnarray}
where $\mu$ is the reference mass scale in the minimal subtraction (MS) scheme. The renormalization constants $Z_g, Z_u$ and $Z_D$
satisfy the relations
\begin{eqnarray}
   Z_g=Z_D^{-3},\hskip1cm Z_u=Z_D^{-1}.
   \label{eq:ren_rel2}
\end{eqnarray}
It should be noted that the graphs corresponding to $\Gamma_{\psi^\dagger\psi\psi}$ and $\Gamma_{\psi^\dagger\psi^\dagger \psi\psi}$
differ only by one external vertex and thus give rise to equal renormalization of the rate constant $\lambda_0 D_0$.
Singularities presented in these (1PI) Green functions may be eliminated by the following renormalization of $\lambda_0$:
\begin{eqnarray}
   \lambda_0 = \lambda \mu^{-2\Delta} Z_D^{-1} Z_\lambda.
   \label{eq:ren_rel3}
\end{eqnarray}
The explicit form of the renormalization constants $Z_D$ and $Z_\lambda$ at the two-loop order can be found
in the appendix.\\
\section{Fixed points}
The coefficient functions of the RG operator
\begin{align}
  & D_{{\rm RG}} = \mu\frac{\partial}{\partial\mu}\biggl|_0 = \mu\frac{\partial}{\partial \mu} +\sum_{g_i}
 \beta_{i}\frac{\partial}{\partial g_i}
  -\gamma_D D\frac{\partial}{\partial_D},
\end{align}
where the bare parameters are denoted with the subscript ``0'', are defined as
\begin{align}
  & \gamma_D=\mu\frac{\partial  \ln Z_D}{\partial_\mu}\biggl|_0, \beta_{i}=\mu\frac{g_i}{\mu}\biggl|_0,
\end{align}
with the charges $g_i=\{g,u,\lambda \}$.
 From this definition and from relations (\ref{eq:ren_rel1}),(\ref{eq:ren_rel2}) and (\ref{eq:ren_rel3}) it follows that
\begin{align}
   &  \beta_g = g(-2\epsilon-\eta+3\gamma_D),\hskip0.5cm \beta_u = u[-\eta+\gamma_D],\hskip0.5cm
   \beta_\lambda = \lambda(2\Delta-\gamma_\lambda+\gamma_D)
   \label{def_beta_func}
\end{align}
The scaling regimes are associated with the fixed points of the corresponding RG functions. The fixed points
 are defined as such points
$g^*,u^*,\lambda^*$ for which the $\beta$ functions vanish
\begin{align}
  \beta_g (g^*,u^*,\lambda^*)=\beta_u(g^*,u^*,\lambda^*)=\beta_\lambda(g^*,u^*,\lambda^*)=0.
  \label{eq:fix_point}
\end{align}
The type of the fixed point is determined by the eigenvalues of the matrix $\Omega=\{\Omega_{ik}=\partial\beta_i/\partial g_k \}$,
where $\beta_i$ is the full set of $\beta$ functions (\ref{def_beta_func}) and $g_k$ is the full set of charges $\{g,u,\lambda\}$ .
The IR asymptotic behavior is governed by the IR stable fixed points, for which all eigenvalues of $\Omega$ matrix are positive. \\
It is easy to see that the functions $\beta_g$ and $\beta_u$ satisfy relation $\beta_g/g - 3\beta_u/u = 2(\eta-\epsilon)$.
This means that they cannot be equal zero simultaneously for the finite
values of the charges $g$ and $u$. The only exception is the instance $\epsilon=\eta$, which should be studied
separately. For general case $\epsilon\neq\eta$ we have to set either $u=0$ or $u=\infty$ and rescale $g$ in such a way, that
$\gamma_D$ remains finite \cite{Antonov}.\\
In what follows we present the results for fixed points, anomalous dimensions and eigenvalues of the $\Omega$ matrix to the second
order of perturbation theory. However, we would like to stress, that the form of $\beta$ functions (\ref{def_beta_func}) allows
to calculate the anomalous dimensions $\gamma_{D}$ and $\gamma_{\lambda}$ exactly (without any second-order correction).\\
In \cite{Adzhemyan2} independence of the renormalization constant $Z_D$ on
the exponents $\eta$ at the two-loop approximation has been conjectured. It implies that we may use the
choice $\eta=0$, which we have applied in our calculations of the renormalization constants $Z_D$
and $Z_\lambda$.
\\
Let us consider the ``rapid-change mode'' ($u\rightarrow\infty$).
It is convenient to introduce new variables $w=1/u, g'=g/u^2$ and the corresponding $\beta$ functions
obtain the form
\begin{align}
   & \beta_{g'} = g'[\eta-\epsilon+\gamma_D],\hskip0.5cm \beta_w = w[\eta-\gamma_D],\hskip0.5cm
    \beta_\lambda = \lambda(2\Delta-\gamma_\lambda+\gamma_D).
\end{align}
The ``rapid-change model'' corresponds to the fixed point with $w^*=0$. In this case four stable IR fixed points are realized:
\begin{align}
    \mbox{FP 1A: } & w^*=0,\mbox{ } {g'}^*=0,\mbox{ } \lambda^* = 0 \nonumber \\
    & \gamma_D =0,\mbox{ } \gamma_\lambda = 0 \\
    & \Omega_1 = \eta-2\epsilon,\mbox{ } \Omega_2=\eta,\mbox{ } \Omega_3 = 2\Delta \nonumber \\
    \mbox{FP 1B: } & w^*=0,\mbox{ } {g'}^*=0,\mbox{ } \lambda^* = -4\pi\Delta \nonumber \\
    & \gamma_D = 0,\mbox{ } \gamma_\lambda = 2\Delta \\
    & \Omega_1 = \eta-2\epsilon,\mbox{ } \Omega_2=\eta,\mbox{ } \Omega_3=-2\Delta \nonumber \\
    \mbox{FP 2A: } & w^*=0,\mbox{ } {g'}^* = 8\pi(2\epsilon-\eta),\mbox{ } \lambda^* = 0 \nonumber \\
    & \gamma_D=2\epsilon-\eta,\mbox{ } \gamma_\lambda=0 \\
    & \Omega_1=2\epsilon-\eta,\mbox{ } \Omega_2=2\eta-2\epsilon,\mbox{ } \Omega_3=2\Delta+2\epsilon-\eta \nonumber \\
   \mbox{FP 2B: } & w^*=0,\mbox{ } {g'}^* = 8\pi(2\epsilon-\eta), \nonumber \\
                  &  \lambda^* = -2\pi(2\Delta+2\epsilon-\eta)-\pi\frac{2\Delta+2\epsilon-\eta}{\epsilon}(2\epsilon-\eta)(\Delta+\epsilon)
 \nonumber \\
   & \gamma_D=2\epsilon-\eta,\mbox{ } \gamma_\lambda=2\Delta+2\epsilon-\eta \\
   & \Omega_1 = 2\epsilon-\eta,\mbox{ } \Omega_2=2\eta-2\epsilon,\mbox{ } \Omega_3=-2\Delta-2\epsilon+\eta. \nonumber
\end{align}
For the analysis of the regime $u\rightarrow 0$ (quenched velocity field)
 we introduce the new variable $g''\equiv g/u^2$. Hence the corresponding $\beta$ functions have the form
\begin{align}
 & \beta_{g''} = g''[-2\epsilon+2\gamma_D],\hskip0.5cm \beta_u = u[-\eta+\gamma_D],\hskip0.5cm
    \beta_\lambda = \lambda(2\Delta-\gamma_\lambda+\gamma_D)
\end{align}
Also in this case there are four possible IR stable fixed points:
\begin{align}
   \mbox{FP 3A: } & u^*=0,\mbox{ } {g''}^*= 0,\mbox{ } \lambda^* = 0 \nonumber \\
   & \gamma_D=0,\mbox{ } \gamma_\lambda=0 \\
   & \Omega_1=-2\epsilon,\mbox{ } \Omega_2=-\eta,\mbox{ } \Omega_3=2\Delta \\
   \mbox{FP 3B: } &  u^*=0,\mbox{ } {g''}^*= 0,\mbox{ } \lambda^*=-4\pi\Delta   \nonumber\\
   & \gamma_D=0,\mbox{ } \gamma_\lambda=2\Delta \\
   & \Omega_1=-2\epsilon,\mbox{ } \Omega_2=-\eta,\mbox{ } \Omega_3=-2\Delta \nonumber \\
   \mbox{FP 4A: } & u^*=0,\mbox{ } {g''}^*=8\pi\epsilon+2\pi(4\Delta+3\epsilon)\epsilon,\mbox{ } \lambda^* = 0  \nonumber\\
   & \gamma_D=\epsilon,\mbox{ } \gamma_\lambda=0 \\
   & \Omega_1=2\epsilon-\frac{4\Delta+3\epsilon}{2}\epsilon,\mbox{ } \Omega_2=-\eta+\epsilon,\mbox{ } \Omega_3 =2\Delta+\epsilon \nonumber \\
   \mbox{FP 4B: } & u^*=0,\mbox{ } {g''}^*=8\pi\epsilon+2\pi(4\Delta+3\epsilon)\epsilon,      \nonumber\\
   &  \lambda^* =  -2\pi(2\Delta+\epsilon) +\pi(2\Delta+\epsilon)({\rm Q}\epsilon-\Delta)\nonumber\\
   & \gamma_D=\epsilon,\mbox{ } \gamma_\lambda=2\Delta+\epsilon \\
   & \Omega_1=2\epsilon-\frac{4\Delta+3\epsilon}{2}\epsilon,\mbox{ } \Omega_2=-\eta+\epsilon,\mbox{ }
\Omega_3 =-2\Delta-\epsilon, \nonumber
\end{align}
where ${\rm Q}=1.159$ is a numerical constant resulting from a simple but cumbersome two-loop integral.\\
In the special case $\epsilon=\eta$ the functions $\beta_g$ and $\beta_u$ become proportional and this leads to the
degeneration of fixed point. Instead of just plain fixed point, we have a line of fixed points in the $g-u$ plane.

\begin{align}
  \mbox{FP 5A: } & \lambda^* = 0 \nonumber\\
  & \gamma_D=\epsilon, \mbox{ }\gamma_\lambda = 0 \\
  & \Omega_1 = G_1(u^*)\epsilon+G_2(u^*)\epsilon, \Omega_2 = 0, \Omega_3 =2\Delta+\epsilon \nonumber\\
  \mbox{FP 5B: } & \lambda^*\neq 0 \nonumber \\
  & \gamma_D=\epsilon,\mbox{ } \gamma_\lambda = 2\Delta+\epsilon \\
  & \Omega_1 = G_1(u^*)\epsilon+G_2(u^*)\epsilon, \Omega_2 = 0, \Omega_3 = -2\Delta-\epsilon \nonumber
\end{align}
where the functions $G_1$ and $G_2$ are given as
\begin{align}
  & G_1(u^*) = \frac{u^*+2}{u^*+1} \nonumber\\
  & G_2(u^*)  = -\frac{1}{4(1+u^*)}\biggl[
  u(3u^3+15u^2+26u+18)\ln\frac{u^2+2u}{(1+u)^2}+\nonumber\\
   & 2(4\xi+3)+3u^2+9u \biggl]
\end{align}
The zero eigenvalue $\Omega_2$ is connected with the existence of a marginal direction in the $g-u$ plane (along the line of the fixed
points).\\
The ``real problem'' corresponds to the value $\epsilon=4/3$, which leads to the famous Kolmogorov ``five-thirds law'' \cite{Frisch} for
the spatial velocity statistics. By direct numerical calculation it may be easily shown that in this case ($\epsilon=4/3$)
the eigenvalue $\Omega_1$ is positive. Of course this holds just for the physically possible values of parameters $u$ and $\xi$.
One has to consider $u>0$ and $\xi<0$, which corresponds to the space dimensions lesser than two.
The conclusion is that all fixed points on the line are IR stable.

\section{Conclusions}
This paper is devoted to the study of reaction kinetics of the annihilation reaction $A+A\rightarrow\varnothing$.
In order to use the technique of the perturbative renormalization a field-theoretic model is constructed.
All the calculations were performed to the second order of the perturbation theory. The IR stable fixed points,
dimensions and corresponding regions of stability of fixed points are calculated.\\
The technically relatively simple model of velocity fluctuations used here is a convenient
starting point for more realistic high-loop calculations based on the stochastic
Navier-Stokes equations. These calculations are almost completed and we hope
to publish them in the near future.\\
 The work was supported by VEGA grant 0173 of Slovak Academy of Sciences, and by Centre of Excellency for
Nanofluid of IEP SAS. This article was also created by implementation
of the Cooperative phenomena and phase transitions
in nanosystems with perspective utilization in nano- and
biotechnology projects No 26220120021 and No 26220120033. Funding for the operational
research and development program was provided by
the European Regional Development Fund.
\section*{Appendix A: Explicit form of renormalization constants $Z_D$ and $Z_\lambda$}
For brevity we use the following normalization of charges
\begin{align}
 & \overline{g}=\frac{g}{4u(1+u)}\frac{S_d}{(2\pi)^d},\hskip0.5cm \overline{\lambda}=\lambda\frac{S_d}{(2\pi)^d}, \hskip0.5cm
  S_d=\frac{2\pi^{d/2}}{\Gamma(\frac{d}{2})}. \nonumber
\end{align}
The following results are presented directly for the space dimension $d=2$
 \begin{align}
    & Z_D = 1 - \frac{ \overline{g}}{2\epsilon} + A(u)\frac{\overline{g}^2}{2\epsilon} -
   \frac{1}{2(1+u)}\frac{\overline{g}^2}{(2\epsilon)^2}\\
    & A(u)=\frac{1}{8(1+u)}\biggl[4\xi+3+u+u(u+2)(u+3)\ln\frac{u^2+2u}{(1+u)^2} \biggl] \nonumber\\
   & Z_\lambda = 1 - \frac{\overline{\lambda}}{2\Delta}  + \frac{\overline{\lambda}^2}{(2\Delta)^2}+
  B(u)\frac{ \overline{g\lambda}}{2(\epsilon-\Delta)}-\frac{\overline{g\lambda}}{4\Delta(\epsilon-\Delta)} \\
  & B(u)= -\frac{1}{2}\biggl[\xi+1+\ln\frac{1+u}{2}+(2u+1)\ln\frac{2u+1}{2u+2} \biggl]
       + \frac{2(1+u)}{\pi}\int_{-1}^1 dz\mbox{ }  F(z)
 \nonumber\\
  & F(z)  = \frac{(1-z^2)^{1/2}}{(1-u)^2+4uz^2} \biggl\{\frac{u-1}{2}\ln\frac{1+u}{2}-\frac{2(1+u)z}{\sqrt{1-z^2}}
 \biggl[\frac{\pi}{2}-\arctan\sqrt{\frac{1+z}{1-z}} \biggl]+\nonumber\\
& \frac{(u+3)z}{\sqrt{2(1+u)-z^2}}\biggl[\pi-\arctan\frac{z+u+1}{\sqrt{2(1+u)-z^2}}-
\arctan\frac{(2+z)}{\sqrt{2(1+u)-z^2}} \biggl]\biggl\}\nonumber
\end{align}

\end{document}